\documentclass[prl,twocolumn,showpacs,amsmath,superscriptaddress,psfig,amssymb]{revtex4-1}

\usepackage{graphicx}
\usepackage{dcolumn}
\usepackage{bm}

\begin{document}


\title{Unsaturated both large positive and negative magnetoresistance in Weyl Semimetal TaP}

\author{Jianhua Du}
\affiliation {Department of Physics, Zhejiang University, Hangzhou 310027, China}

\author{Hangdong Wang}
\affiliation {Department of Physics, Hangzhou Normal University, Hangzhou 310036, China}

\author{Qianhui Mao}
\affiliation {Department of Physics, Zhejiang University, Hangzhou 310027, China}

\author{Rajwali Khan}
\affiliation {Department of Physics, Zhejiang University, Hangzhou 310027, China}

\author{Binjie Xu}
\affiliation {Department of Physics, Zhejiang University, Hangzhou 310027, China}

\author{Yuxing Zhou}
\affiliation {Department of Physics, Zhejiang University, Hangzhou 310027, China}

\author{Yannan Zhang}
\affiliation {Department of Physics, Zhejiang University, Hangzhou 310027, China}

\author{Jinhu Yang}
\affiliation {Department of Physics, Hangzhou Normal University, Hangzhou 310036, China}

\author{Bin Chen}
\affiliation {Department of Physics, Hangzhou Normal University, Hangzhou 310036, China}

\author{Chunmu Feng}
\affiliation {Department of Physics, Zhejiang University, Hangzhou 310027, China}

\author{Minghu Fang}
\email{mhfang@zju.edu.cn}
\affiliation {Department of Physics, Zhejiang University, Hangzhou 310027, China}
\affiliation {Collaborative Innovation Center of Advanced Microstructures, Nanjing 210093, China}

\date{\today}

\begin{abstract}
\noindent After growing successfully TaP single crystal, we measured its longitudinal resistivity ($\rho_{xx}$) and Hall resistivity ($\rho_{yx}$) at magnetic fields up to 9T in the temperature range of 2-300K. It was found that at 2K its magnetoresistivity (MR) reaches to 3.28$\times$10$^5$$\%$, at 300K to 176$\%$ at 8T, and both do not appear saturation. We confirmed that TaP is indeed a low carrier concentration, hole-electron compensated semimetal, with a high mobility of hole $\mu_h$=3.71$\times$$10^5$ $cm^2$/V s, and found that a magnetic-field-induced metal-insulator transition occurs at room temperature. Remarkably, as a magnetic field (\textit{H}) is applied in parallel to the electric field (\textit{E}), the negative MR due to chiral anomaly is observed, and reaches to -3000$\%$ at 9T without any signature of saturation, too, which distinguishes with other Weyl semimetals (WSMs). The analysis on the Shubnikov-de Haas (SdH) oscillations superimposing on the MR reveals that a nontrivial Berry's phase with strong offset of 0.3958 realizes in TaP, which is the characteristic feature of the charge carriers enclosing a Weyl nodes. These results indicate that TaP is a promising candidate not only for revealing fundamental physics of the WSM state but also for some novel applications.

\end{abstract}

\pacs{72.90+y;72.10.-d;72.15.Om}
\maketitle

The Weyl semimetal (WSM) phase, in which the bulk electronic bands disperse linearly along all momentum direction through a node, the Weyl point, in a three-dimension (3D) analogue of graphere, can be viewed as an intermediate phase between a trivial insulator and a topological insulator\cite{Murakami 2007,Balents 2011,Burkov 2011-1,Burkov 2011-2,Halasz 2012}. Although a number of candidates for a WSM were previously proposed, such as, Y$_2$Ir$_2$O$_7$\cite{Wan 2011} and HgCr$_2$Se$_4$\cite{Xu 2011} compounds, in which magnetic order breaks the time-reversal symmetry, and LaBi$_{1-x}$Sb$_x$Te$_3$\cite{Liu 2014} compound, in which fine-tuning the chemical composition is necessary for breaking inversion symmetry, a WSM has not realized experimentally in any of these compounds due to either no enough large magnetic domain or difficulty to tune the chemical composition within 5$\%$. Very recently, the theoretical proposal\cite{Weng 2015,Huang 2015} for a WSM in a class of stoichiometric materials, including TaAs, TaP, NbAs and NbP, which break crystalline inversion symmetry, has been soon confirmed by the experiments\cite{Lv 2015,Shekhar 2015,Huang 2015-1,Zhang 2015}, except for TaP due to difficulty to grow large crystal. The exotic transport properties exhibiting in these materials ignite an extensive interesting in both the condensed matter physics and material science community, especial for their extremely large magnetoresistance (MR) and ultrahigh mobility of charge carriers.

Materials with large MR have been used as magnetic sensors\cite{Lenz 1990}, in magnetic memory\cite{Moritomo 1996}, and in hard drives\cite{Daughton 1999} at room temperature. Large MR is an uncommon property, mostly of magnetic compounds, such as a giant magnetoresistance (GMR)\cite{Egelhoff 1995} emerging in Fe/Cr thin-film, and colossal magnetoresistance (CMR) in the manganese based perovskites\cite{Ramirez 1997,Jin 1994}. In contrast, ordinary MR, a relatively weak effect, is commonly found in non-magnetic compounds and elements\cite{Yang 1999}. Magnetic materials typically have negative MR. Positive MR is seen in metals, usually at the level of a few percent, and in some semiconductors, such as 200$\%$ at room temperature in Ag$_{2+\delta}$(Te,Se)\cite{Xu 1997}, comparable with those of materials showing CMR\cite{Ishiwata 2013}, and semimetals, such as high-purity bismuth, graphite\cite{Solin 2000}, and 4.5$\times$10$^4$$\%$ in WTe$_2$\cite{Ali 2014}. In the semimetals, very high MR is attributed to a balanced hole-electron "resonance" condition, as described in Ref.\cite{Ali 2014}. WSM provides another possibility to realize extremely large MR, as observed in TaAs\cite{Huang 2015-1,Zhang 2015}, NaAs\cite{Yang 2015} and NbP\cite{Shekhar 2015}, moreover, the helical Weyl fermions with unprecedented mobility are well protected from defect scattering by real spin conservation associated to the chiral Weyl nodes. Thus, it is very interesting not only for fundamental physics, but also for practical application to search for more WSM compounds and characterize their MR behavior.

TaP crystalizes in a body-centered tetragonal lattice with nonsymmorphic space group $\textit{I4$_1$md} (No. 109)$, which lacks inversion symmetry, shown in Fig. 1(a), which is one of the member predicted as a WSM in Refs.\cite{Weng 2015,Huang 2015}. In this letter, we grew successfully TaP single crystal, then measured its longitudinal resistivity ($\rho_{xx}$) and Hall resistivity ($\rho_{yx}$) at magnetic fields up to 9T and in the temperature range of 2-300K. It is found that at 2K its MR reaches to 3.28$\times$10$^5$$\%$, and at 300K to 176$\%$ at 8T, and both do not appear saturation, and a magnetic-field-induced metal-insulator transition occurs at room temperature. Hall resistivity measurements confirmed that TaP is indeed a low carrier concentration, hole-electron compensated semimetal, with the high mobility $\mu_h$=3.71$\times$$10^5$ $cm^2$/V s. Remarkably, as a magnetic field (\textit{H}) is applied in parallel to the electric field (\textit{E}), the negative MR due to chiral anomaly is observed, reaches to near -3000$\%$ at 9T without any signature of saturation, which distinguishes with other WSMs and may originate from a larger distance between Weyl nodes in the momentum space. We have also confirmed that a nontrivial Berry's phase with strong offset of 0.3958 realizes in TaP, which is the characteristic feature of the charge carriers enclosing a Weyl nodes.

\begin{figure}
  \includegraphics[width=8cm]{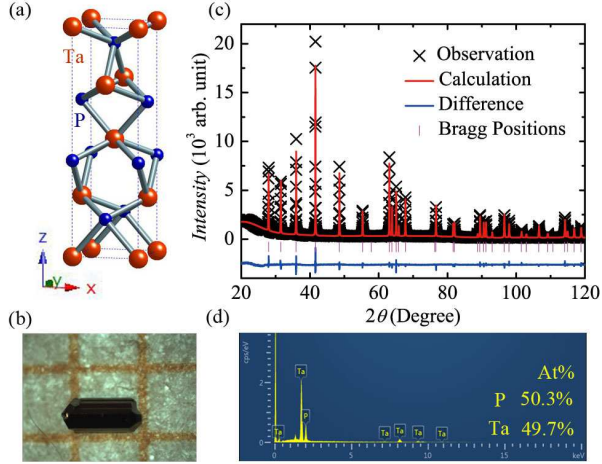}\\
  \caption{(Color online) (a) Crystal structure of TaP with a body-ceentered tetragonal structure. (b) Photo of TaP crystal. (c) XRD pattern of powder obtained by grinding TaP crystals. Its Rietveld refinement is shown by the line. (d) Energy Dispersive X-ray Spectrometer (EDXS) for the TaP crystal.}\label{}
\end{figure}

Single crystals of TaP were grown using a chemical vapor transport method. Polycrystalline TaP prepared previously was filled in the quartz tube with a 10mg/cm$^3$ Iodine as a transporting agent. After evacuated and sealed, the quartz tube was heated for three weeks in a tube furnace with a temperature field of $\Delta$T=950-850$^o$C. Large polyhedral crystals with dimensions up to 1.0 mm were obtained, as shown in Fig. 1(b). Energy Dispersive X-ray Spectrometer (EDXS) was used to determine the crystal composition, and stoichiometric TaP was confirmed. X-ray diffraction (XRD) pattern [see Fig. 1(c)] at room temperature of the TaP powder by grinding pieces of crystals confirms its tetragonal structure, and its Rietveld refinement (reliability factor: $R_{wp}$= 7.18$\%$, $\chi^2$= 2.973) gives the lattice parameters of \textit{a} = 3.3184($\pm$0.0001) {\AA} and \textit{c} = 11.3388($\pm$0.0001) {\AA}. Electrical resistivity in magnetic field (\textit{H}) and Hall resistivity measurements were carried out using a Quantum Design Physical Property Measurement System (PPMS).

\begin{figure}
  \includegraphics[width=8cm]{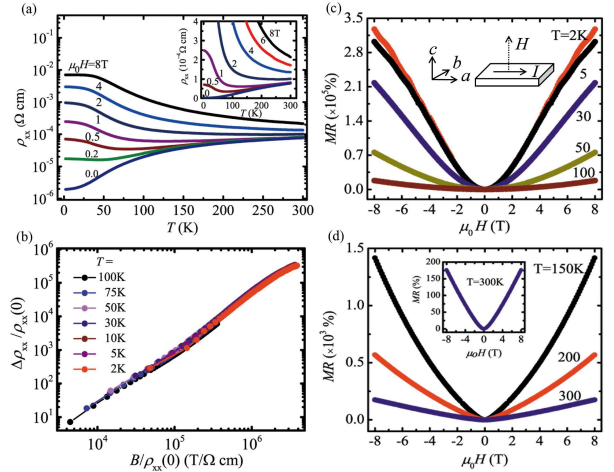}\\
  \caption{(Color online)(a) Temperature dependence of longitudinal resistivity, $\rho$$_{xx}$(T), measured at different magnetic field for TaP single crystal. (b) MR scaling law with magnetic field (c) Magnetic field dependence of MR for below 100K. (d) for above 100K.}\label{}
\end{figure}

Figure 2 summarizes the resistivity ($\rho_{xx}$) results measured at different \textit{H}, which was applied in \textit{c} axis, and normal to the current, and various temperatures (\textit{T}) of TaP crystal. As shown in Fig. 2(a), the temperature dependence of resistivity, $\rho_{xx}$(T), at \textit{H}=0T displays a metallic behavior with $\rho$(300K)=77.98 $\mu\Omega$ cm, and $\rho$(2K)=1.933 $\mu\Omega$ cm, thus estimated residual resistivity ratio [RRR] =$\rho$(300K)/$\rho$(2K)$\sim$ 40. It is obvious that the crossover from "metallic" to "insulating" behavior as a magnetic field is applied, that looks like a magnetic-field-induced metal-insulator transition. On lowering the temperature the resistivity increases, as it does in an insulator, but then saturates towards field-dependent constant values at the lowest temperature, more clearly seen in the log-log plot in the main panel of Fig. 2(a). The similar behavior has also been observed in the conventional semimetals, such as high-purity Bi\cite{Yang 1999}, graphite\cite{Kopelevich 1999,Kopelevich 2003}, WTe$_2$\cite{Ali 2014} and $\beta$-Ag$_2$Te\cite{Xu 1997,Zhang 2011}, and in the Dirac semimetal Cd$_3$As$_2$\cite{He 2014} and other Weyl semimetals TaAs\cite{Zhang 2015,Huang 2015}, NbAs\cite{Ghimire 2015} and NbP\cite{Shekhar 2015,Wang 2015}. This similarities invite an interpretation that ascribes this interesting behavior to the properties shared by these compounds, namely, low carrier density and an equal number of electrons and holes (compensation). In contrast with that the metallic behavior remains at higher temperature even at high field in the WTe$_2$\cite{Ali 2014} and $\beta$-Ag$_2$Te\cite{Xu 1997,Zhang 2011}, the metal-insulator transition occurs at room temperature at above 2T field in the semimetal TaP, as shown in the inset in Fig.1(a), as well as TaAs\cite{Zhang 2015}.

The relative change of the MR at various temperatures as a function of magnetic field is plotted in Fig.2(c) and (d) by the typical definition of MR\cite{Pippard 1989}:$\frac{\Delta\rho}{\rho(0)}$=[$\frac{\rho(H)-\rho(0)}{\rho(0)}$]$\times$100$\%$, where $\rho(H)$ is the resistivity measured at \textit{H} for each given isotherm. At 2K, the MR reaches to 3.28$\times$10$^5$$\%$ at 8T, two times larger than that in the recently reported compound WTe$_2$ at 2K in 8T although its RRR is 1-2 orders of magnitude less than that of the latter, and does not appear saturation in the highest field in our measurements. Remarkable, the MR at room temperature (300K) can reach to 176$\%$ at 8T, and increases almost linearly with field without any saturation, indicating that these WSMs can be used in the magnetic device in the future. It can be seen explicitly that there is a transition from quadratic MR at low magnetic fields to linear MR at high magnetic fields at a given temperature, as expected theoretically for the WSMs\cite{Ramakrishnan 2015}. The relative MR of many metal and semimetals can be represented by the form, commonly referred to as Kohler's rule\cite{Pippard 1989}, $\Delta\rho/\rho(0)$=\textit{F}[\textit{H}/$\rho(0)$], where \textit{F}(\textit{H}) usually follows a power law. As shown in Fig.2(b), although all the MR data below 100K collapse onto the same curve, the MR does not follow a simple power law. And we found that the MR data above 100K does not collapse this curve, indicating that phonon scattering plays a role at higher temperatures. The following Hall resistivity and SdH oscillation analysis help to understand this complexity. Another, there are Shubnikov-de Haas (SdH) oscillations superimposing on the MR, which amplitude increases with decreasing temperature. Detailed analysis of the SdH data will be presented below.

\begin{figure}
  \includegraphics[width=8cm]{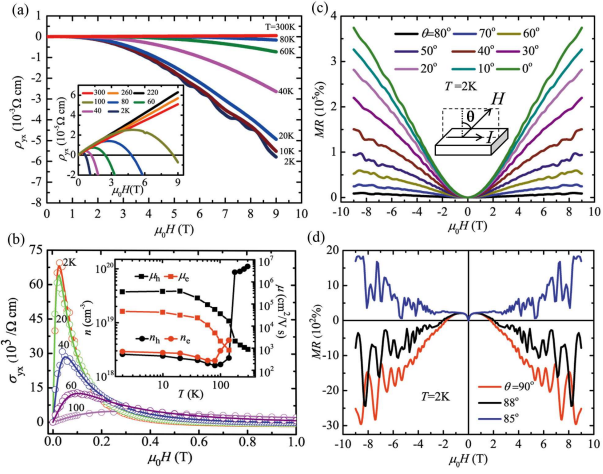}\\
    \caption{(Color online) (a) The Hall resistivity versus magnetic field from 2 to 300K, strong SdH oscillation were observed below 10K. Inset: Hall resistivity for the higher temperatures and for lower temperatures in lower fields. (b) Hall conductivity $\sigma_{yx}$ at five representative temperatures as a function of field. The solid lines are the fitting curves from the two-carrier model. Inset: The temperature dependence of the mobilities and carrier concentrations of the electrons and holes. (c) The MR versus magnetic field measured at 2K for various \textit{H} alignment with respect to \textit{I}. (d) The MR versus magnetic field measured at 2K for \textit{H} near in parallel to \textit{I}. }\label{}
\end{figure}

\begin{figure}
  \includegraphics[width=8cm]{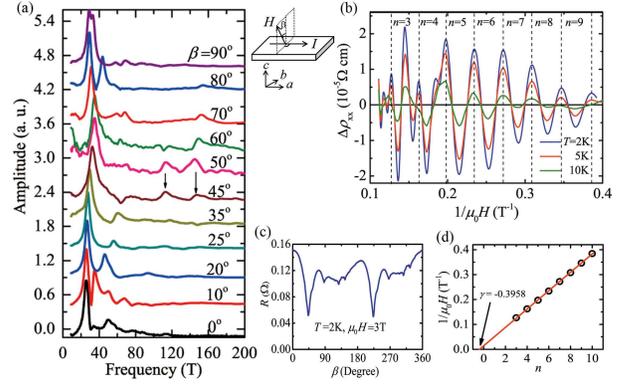}\\
  \caption{(Color online) (a) The Fourier transformation for magnetoresistivity after subtracting the non-oscillation part, measured at 2K for various \textit{H} alignment with respect to \textit{c} axis, and keeping $H \perp I$. (b) $\Delta\rho_{xx}$ as a function of 1/$\mu_0H$ at 2K, 5K, and 10K, where $\Delta\rho_{xx}$ is the oscillation part in the magetoresistivity $\rho_{xx}$ after subtracting $\rho_0$ for $\beta$=0. (c) The angle dependence of resistance , $R_{xx}$($\beta$), measured at 2 K and 3 T. (d) Linear fitting of the Landau fan diagram for $F_1$= 25.6 T, showing a non-trivial Berry's phase with strong offset of 0.3958. }\label{}
\end{figure}

Then, we discuss the charge carrier information about TaP compound. Figure 3(a) shows the Hall resistivity as a function of magnetic field, $\rho$$_{yx}$(\textit{H}), measured at various temperatures. At higher temperatures (above 140K), the positive slope of $\rho$$_{yx}$(\textit{H}) indicates that the holes dominate the main transport processes. At lower temperatures (below 140K), even at the lowest temperature (2K) the slope of $\rho$$_{yx}$(\textit{H}) changes from a positive to a negative value with increasing field, as shown in the inset of Fig. 3(a), indicating the coexistence both electron and hole carrier charges. SdH oscillations superimposed on the $\rho$$_{yx}$(\textit{H}) at higher field below 10K is also obvious. As done in Ref.\cite{Zhang 2015} for the TaAs compound, we fitted the Hall conductivity tensor calculated using $\sigma_{xy}$=$\rho_{yx}$/($\rho_{xx}^2$+$\rho_{yx}^2$) by adopting a two-carrier model\cite{Hurd 1972}, $\sigma_{xy}$=[$n_h$$\mu_h^2$$\frac{1}{1+(\mu_hH)^2}$-$n_e$$\mu_e^2$$\frac{1}{1+(\mu_eH)^2}$]. Where $n_e$ ($n_h$) and $\mu_e$ ($\mu_h$) denote the carrier concentrations and mobilities of the electrons (holes), respectively. Figure 3(b) presents the fitting for $\sigma_{xy}$ data measured at five representative temperatures. It should be pointed out that the Hall conductivity above 180K is mainly from the hole band due to the thermal depopulation of hole-like pockets just above the Fermi level, thus the $n_h$ and $\mu_h$ values above 180K were estimated by using a single band fitting. The obtained $n_e$ ($n_h$) and $\mu_e$ ($\mu_h$) values by fittings in whole temperature range are shown in the inset of Fig.3(b). At first, it is obvious that the $n_e$ and $n_h$ values at lower temperatures (below 100K) are almost the same, such as at 2K, $n_e$=2.58$\times$$10^{18}$/cm$^3$, $n_h$=2.90$\times$$10^{18}$/cm$^3$, indicating TaP is indeed a low carrier concentration, hole-electron compensated semimetal, which is in consistent with the discussions above about the extremely large MR. Notes that the $n_h$ has an obvious increase at 140K with increasing temperature. Secondly, at lower temperatures, the mobility of hole, $\mu_h$, is one order of magnitude larger than $\mu_e$, such as at 20K, $\mu_h$=3.71$\times$$10^5$ $cm^2$/V s, and $\mu_e$=3.04$\times$$10^4$ $cm^2$/V s, in contrast with that in other WSMs, such as TaAs\cite{Zhang 2015,Huang 2015}, NbAs\cite{Ghimire 2015} and NbP\cite{Shekhar 2015}, in which usually $\mu_e$$>$$\mu_h$. Note that as increasing temperature, both $\mu_h$ and $\mu_e$ have an obvious decrease around 140K due to enhancement of phonon thermal scattering. The mobility of carriers in all the WSMs, such as TaAs, NbAs and NbP, verified recently, as well as TaP reported here, are comparable, but much higher than that in the conventional semimetals, such as WTe$_2$\cite{Ali 2014} and high purity Bi\cite{Yang 1999}, graphite\cite{Kopelevich 1999,Kopelevich 2003}. This is related to the existence of Weyl fermion in these compounds.

In the WSM, due to the presence of chiral Weyl node pairs, unconventional negative MR induced by the Adler-Bell-Jackiw anomaly\cite{Adler 1969,Bell 1969} (also named as the chiral anomaly) would be expected when \textit{H} is applied in parallel to the electric field \textit{E}, \textit{i.e.}, the current direction \textit{I}. Thus, we measured the MR at 2K for the different angles, $\theta$, denoted as in the inset of Fig.3(c), between \textit{H} and \textit{I}. As shown in Fig. 3(d), when $\theta$=90$^o$, \textit{i.e.}, \textit{H} $\parallel$ \textit{I}, the negative MR is indeed observed, reaches to near -3000$\%$ at 9T, and exhibits no signature of saturation even at the highest field in our measurements. This behavior is different with that in other WSMs, such as in TaAs\cite{Huang 2015-1}, and NbP\cite{Wang 2015}, in which MR changes soon from negative to positive value as \textit{H} increases, and may be due to TaP having a larger distance of Weyl nodes in the momentum space\cite{Xu 2015-1}. Another, the MR curve has a small positive peak below 2T, which was also observed in Bi$_{1-x}$Sb$_x$\cite{Kim 2013}, TaAs\cite{Huang 2015-1,Zhang 2015} and NbP\cite{Wang 2015}, and attributed to a weak anti-localization. It is obvious that the SdH oscillation superimposing onto the MR. If $\theta$ decreases to 88$^o$, the unsaturated negative MR remains, but its value decreases. When $\vartheta$ $\leq$ 85$^o$, the MR becomes positive value and reaches the maximum value as $\theta$ = 0$^o$, as shown in Fig.(c).

In order to get the information of the Fermi surface of TaP, we also measured the magnetoresistivity, $\rho_{xx}$(\textit{H}) (not shown here) for different angles $\beta$, between \textit{H} and \textit{c} axis, represented in the inset of Fig.4, and keeping \textit{H} $\perp$ \textit{I}. Figure 4(c) shows the angle dependence of resistance measured at 2K and 3.0T. It is found that the resistance in magnetic field exhibits maximum at $\beta$=0, and reaches a minimum value at $\beta$=45$^o$, indicating the anisotropy of the Fermi surface topology. In fact, for all the angles, strong Shubnikov-de Haas (SdH) oscillations superimpose onto the $\rho_{xx}$(H). 

First, we discuss another interesting consequence due to the presence of chiral Weyl node pairs: a nontrivial Berry's phase ($\Phi_B$) realizes in TaP, which is the characteristic feature of the charge carriers that have \textit{k}-space cyclotron orbits enclosing a Dirac points.\cite{Mikitik 1999,Mikitik 2004,Novoselov 2005,Zhang 2005}. We use the expression\cite{Murakawa 2013}: $\rho_{xx}$=$\rho_0$+$\Delta\rho_{xx}$=$\rho_0$[1+\textit{A}(\textit{B},\textit{T})cos2$\pi$(\textit{F/B}+$\gamma$)], to analyze the SdH oscillations in $\rho_{xx}$(\textit{H}), as \textit{H} $\parallel$ \textit{c} axis, \textit{i.e.} $\beta$=0$^o$ in the Fig. 4. Here $\rho_0$ is the non-oscillatory part of the resistivity, \textit{A}(\textit{B,T}) is the amplitude of SdH oscillations, \textit{B}=$\mu_0$H is the magnetic field, and $\gamma$ is the Onsager phase, and \textit{F}=$\frac{\hbar}{2e\pi}$$A_F$ is the frequency of the oscillations, where $A_F$ is the extremal cross-sectional area of the Fermi surface (FS) associated with the Landau level index \textit{n}, \textit{e} is the elementary charge, and $\hbar$=$\frac{h}{2\pi}$, \textit{h} is the Plank's constant. Figure 4(b) shows the $\Delta\rho_{xx}$ as a function of 1/$\mu_0H$ at 2K, 5K, and 10K after subtracting $\rho_0$. Obviously, there are two sets of oscillations, corresponding to two frequencies of $F_1$= 25.6 T, and $F_1$= 49.6 T, confirmed by its Fourier transformation ($\beta$=0) in Fig.4(a). According to Onsager relation, the extremal cross-sectional areas of the FS were estimated as $A_{F}$=2.439$\times$10$^{-3}$ ${\AA}^2$, 4.725$\times$10$^{-3}$ ${\AA}^2$ corresponding to $F_1$ and $F_2$, respectively, which are very small compared with the whole cross-sectional area of the first BZ\cite{Huang 2015-1}. Then we focus on the SdH oscillation with $F_1$. To determine the Berry's phase of the carriers, we have plotted the Landau fan diagram, as shown in Fig. 4(d), linear fitting gives an Onsager phase of $\gamma$ =-0.3958. The existence of a nontrivial Berry's phase with strong offset of 0.3958 confirms its Weyl fermions behavior. It has been predicted\cite{Huang 2015} by the \textit{ab initio} calculations that there are 12 pairs of Weyl nodes in the Brillouin zone (BZ), the eight pairs locates at $k_z$ $\sim$ $\pm$ 0.6$\pi/c'$ (named W1) where $c'=c$/2 and the other four pairs sit in the $k_z$=0 plane (W2). As \textit{H} is applied along \textit{c} axis, possible WSM electron pockets and hole pockets are quantized equivalently duo to the four-fold rotation symmetry of the Brillouin zone (BZ)\cite{Weng 2015}. So, we expect that the two major oscillation frequencies at $\beta$=0$^o$ are assigned to these two types Weyl nodes.

Second, we focus on the angle ($\beta$) dependence of the SdH oscillations frequency. As shown in Fig.4(a), with increasing $\beta$, $F_1$ peak at $\beta$=0 shifts to a little higher frequency then turns back, exhibiting a near isotropic behavior, as expected for W1 cones\cite{Weng 2015}. But for $F_2$ at $\beta$=0, corresponding to the anisotropic W2 nodes, it is difficult to give clear trend to change with angle $\beta$. Interestingly, there are another two peaks at $F'$=112 T and 146 T for $\beta$= 45$^o$ and 50$^o$ curves, except for the major peak at 32.8T corresponding to W1 nodes. These high frequency peaks may originate from the other massive bands near FS, which seems to provide an explanation for the minimum of MR observed at $\beta$=45$^o$.

In summary, we have measured the longitudinal resistivity ($\rho_{xx}$) and Hall resistivity ($\rho_{yx}$) for TaP single crystal. It is found that at 2K its MR reaches to 3.28$\times$10$^5$$\%$, especially at 300K can reaches to 176$\%$ at 8T, and both do not appear saturation even at the highest field in our measurements. Hall resistivity measurements confirmed that TaP is indeed a low carrier concentration, hole-electron compensated semimetal, with the high mobility. Remarkably, as a magnetic field (\textit{H}) is applied in parallel to the electric field (\textit{E}), the negative MR due to chiral anomaly is observed, reaches to near -3000$\%$ at 9T, and exhibits also no signature of saturation, which distinguishes with other WSMs. Based on the analysis Shubnikov-de Haas (SdH) oscillations superimposing on the MR, it is found that a nontrivial Berry's phase with strong offset of 0.3958 realizes in this compound, which confirms its Weyl fermions behavior.

This work is supported by the National Basic Research Program of China (973 Program) under grant No. 2015CB921004, 2012CB821404, and 2011CBA00103, the Nature Science Foundation of China (Grant No. 11374261, and 11204059) and Zhejiang Provincial Natural Science Foundation of China (Grant No. LQ12A04007), and the Fundamental Research Funds for the Central Universities of China.

During preparation of this manuscript the authors became aware of a related work by C. Shekhar et al. posted on arXiv [arXiv. 1506.06577].


\begin{thebibliography}{apssamp}

\bibitem{Murakami 2007} S. Murakami, New J. Phys. \textbf{9},356(2007)

\bibitem{Balents 2011} L. Balents, Physics \textbf{4}, 36(2011)

\bibitem{Burkov 2011-1} A. A. Burkov and L. Balents, Phys. Rev. Lett. \textbf{107}, 127205(2011)

\bibitem{Burkov 2011-2} A. A. Burkov, M. D. Hook and L. Balents, Phys. Rev. B. \textbf{84}, 235126(2011)

\bibitem{Halasz 2012} G. B. Halasz and L. Balents, Phys. Rev. B. \textbf{85}, 035103(2012)

\bibitem{Wan 2011} X. Wan \textit{et al.} Phys. Rev. B. \textbf{83}, 205101(2011)

\bibitem{Xu 2011} G. Xu,\textit{et al.} Phys. Rev. Lett. \textbf{107}, 186806(2011)

\bibitem{Liu 2014} J. Liu \textit{et al.} Phys. Rev. B. \textbf{90}, 155316(2014)

\bibitem{Weng 2015} Hongming Weng, Chen Fang, Zhong Fang, B. Andrei Bernevig, and Xi Dai, Phys. Rev. X. \textbf{5}, 011029(2015)

\bibitem{Huang 2015} Shin-Ming Huang, \textit{et al.}, arXiv: 1501.00755

\bibitem{Lv 2015} B. Q. Lv \textit{et al.}, arXiv:1503.09188

\bibitem{Shekhar 2015} Chandra Shekhar \textit{et al.}, arXiv:1502.04361

\bibitem{Huang 2015-1} Xiaochun Huang \textit{et al.}, arXiv:1502.01304

\bibitem{Zhang 2015} Chenglong Zhang \textit{et al.}, arXiv:1502.00251

\bibitem{Wang 2015} Zhen Wang \textit{et al.}, arXiv:1506.02282

\bibitem{Lenz 1990} J. E. Lenz, Proc. IEEE \textbf{78} 973(1990)

\bibitem{Moritomo 1996} Y. Moritomo, A. Asamitsu, H. Kuwahara and Y. Tokura, Nature \textbf{380} 141(1996)

\bibitem{Daughton 1999} J. Daughton, J. Magn. Magn. Mater. \textbf{192} 334(1999)

\bibitem{Egelhoff 1995} W. F. Egelhoff \textit{et al.}, J. Appl. Phys. \textbf{78} 273(1995)

\bibitem{Ramirez 1997} A. P. Ramirez, R. J. Cava and J. Krajewski, Nature \textbf{386} 156(1997)

\bibitem{Jin 1994} S. Jin, M. McCormack, T. H. Tiefel and R. Ramesh, J. Appl. Phys. \textbf{76} 6929(1994)

\bibitem{Yang 1999} F. Y. Yang et al. Science \textbf{284} 1335(1999)

\bibitem{Xu 1997} R. Xu, \textit{et al.}, Nature \textbf{390} 57(1997)

\bibitem{Ishiwata 2013} S. Ishiwata et al. Nature Mater. \textbf{12} 512(2013)

\bibitem{Solin 2000} S. A. Solin, T. Thio, D. R. Hines, and J. J. Heremans, Science \textbf{289} 1530(2000)

\bibitem{Ali 2014} Mazhar N. Ali \textit{et al.}, Nature \textbf{514}, 205(2014)

\bibitem{Yang 2015} Xiaojun Yang \textit{et al.}, arXiv:1503.07571(2015)

\bibitem{Kopelevich 1999} Y. Kopelevich \textit{et al.} Phys. Solid State \textbf{41}, 1959(1999)

\bibitem{Kopelevich 2003} Y. Kopelevich \textit{et al.}, Phys. Rev. Lett. \textbf{90}, 156402(2003)

\bibitem{Xu 1997} R. Xu \textit{et al.}, Nature (London) \textbf{390}, 57(1997)

\bibitem{Zhang 2011} W. Zhang \textit{et al.}, Phys. Rev. Lett. \textbf{106}, 156808(2011)

\bibitem{He 2014} L. P. He \textit{et al.}, Phys. Rev. Lett. \textbf{113}, 246402(2014)

\bibitem{Ghimire 2015} N. J. Ghimire \textit{et al.}, arXiv: 1503.07571

\bibitem{Pippard 1989} A. B. Pippard, \textit{Magnetoresitance in Metals}(Cambridge University Press, Cambridge, 1989)

\bibitem{Ramakrishnan 2015} N. Ramakrishnan, M. Milletari, and Shaffique Adam, arXiv: 1501.03815

\bibitem{Hurd 1972} Colin M. Hurd, \textit{The Hall effect in metals and alloys}(Cambridge University Press, Cambridge, 1972)

\bibitem{Murakawa 2013} H. Murakawa \textit{et al.}, Science \textbf{342}, 1490(2013)

\bibitem{Adler 1969} S. Adler, Phys. Rev. \textbf{177}, 2426(1969)

\bibitem{Bell 1969} J. S. Bell and R. Jackiw, Nuovo Cimento A \textbf{60}, 4(1969)

\bibitem{Xu 2015-1} N. Xu \textit{et al.}, arXiv: 1507.03983

\bibitem{Mikitik 1999} G. P. Mikitik, Yu V. Sharlai, Phys. Rev. Lett. \textbf{82}, 2147(1999)

\bibitem{Mikitik 2004} G. P. Mikitik, Yu V. Sharlai, Phys. Rev. Lett. \textbf{93}, 106403(2004)

\bibitem{Novoselov 2005} K. S. Novoselov \textit{et al.}, Nature \textbf{438}, 197(2005)

\bibitem{Zhang 2005} Y. Zhang, Y. W. Tan, H. L. Stormer, P. Kim, Nature \textbf{438}, 201(2005)

\bibitem{Kim 2013} H. J. Kim \textit{et al.}, Phys. Rev. Lett. \textbf{111}, 246603(2013)


\end{thebibliography}
\end{document}